\documentclass[aps,twocolumn,epsfig,graphics,showpacs,floatfix,mathbbm]{revtex4}

\usepackage{amsmath,amsfonts,amssymb,graphics,graphicx,epsfig,color,times,bbm}

\newcommand{\cc}{\mathbbm{C}}
\newcommand{\nn}{\mathbbm{N}}
\newcommand{\rr}{\mathbbm{R}}
\newcommand{\id}{\mathbbm{1}}

\begin{document}

\title{Single-copy entanglement in critical quantum spin chains}

\author{
J.\ Eisert$^{1,2}$ and M.\ Cramer$^3$}

\affiliation{
%\address{
1 QOLS, Blackett Laboratory, 
Imperial College London,
Prince Consort Road, London SW7 2BW, UK\\
2 Institute for Mathematical Sciences, Imperial College London,
Exhibition Rd, London SW7 2BW, UK\\
3 
Institut f{\"u}r Physik, 
Universit{\"a}t Potsdam,
Am Neuen Palais 10, D-14469 Potsdam, Germany
}
\date\today

%\maketitle

\begin{abstract}
We introduce the single-copy entanglement 
as a quantity to assess quantum correlations in the ground state
 in quantum many-body systems. We show for a large class of models
that already on the level of single specimens of spin 
chains, criticality is
accompanied with the possibility of distilling a maximally 
entangled state of arbitrary dimension
from a sufficiently large block deterministically,
 with local operations and classical communication. 
These analytical results -- which refine previous results on the 
divergence of block entropy as the rate at which 
EPR pairs can be distilled from many identically prepared chains, and which 
apply to single systems as encountered in actual experimental situations -- 
are made quantitative for general isotropic translationally 
invariant spin chains that can be mapped onto a quasi-free
fermionic system, and for the anisotropic XY model. For the XX model, we
provide the asymptotic scaling of $\sim (1/6)\log_2( L) $, and
contrast it with the block entropy. The role of superselection rules
on single-copy entanglement in systems consisting of
indistinguishable particles is emphasized.
\end{abstract}
\pacs{03.75.Ss, 03.75.Lm, 03.75.Kk}

\maketitle

%\begin{multicols}{2}

Quantum phase transitions of second order are accompanied with
a divergent length scale: this is the classical correlation
length, the characteristic length associated with
the two-point correlation function
\cite{Sachdev}.
Recently, it has increasingly become clear that one
should expect additional
insight in the scaling of quantum correlations present
in the ground state of a many-body
system at or close to a quantum phase transition
by expressing them in terms of entanglement properties
\cite{Osterloh,Harmonic,Area,Cardy,Latorre1,Korepin1,Keating,GraphStates,Wolf,Orus}.
Entanglement, after all,
plays a fundamental role in quantum phase
transitions at zero temperature.
The theory
of entanglement  in turn  -- developed in the quantum information 
context 
 --  provides tools to 
characterize and quantify 
genuine quantum correlations in contrast to correlations 
that occur in states that can be 
prepared with mere local preparations
together with classical communication (LOCC).
In particular, one finds that in one-dimensional
non-critical harmonic \cite{Harmonic,Area,Cardy}
or quantum spin systems 
\cite{Cardy,Latorre1,Korepin1,Keating,GraphStates},
the degree of entanglement of a block of $L$ systems,
quantified in terms of the entropy of the reduction, typically
saturates 
for large block size, with 
higher-dimensional ``entropy-area laws'' \cite{Area}. 
In contrast, in
critical spin systems or in fermionic systems, 
the entropy of a reduction
has logarithmic corrections
as $L\rightarrow \infty$ \cite{Latorre1,Korepin1,Keating,Wolf}.
These findings are consistent with expectations from conformal
field theory \cite{Cardy}.
Such a behavior of the block entropy has also been related to
the performance of DMRG simulations of ground state properties.
This von-Neumann entropy of a block quantifies the rate at
which one can asymptotically
distill maximally entangled qubit pairs 
under LOCC, when one
has infinitely many identically prepared many-body systems at hand 
\cite{Bennett}.

Yet, in several contexts, in particular for condensed-matter systems,
this asymptotic notion of entanglement 
implicitly referring to joint operations on many identical
systems 
may not always be the most appropriate one. Instead, one may ask:
{\em does a single
specimen of a critical infinite system already contain an
infinite amount of entanglement}? This will be the central question 
of this paper. We introduce the single-copy entanglement to
quantify the quantum correlations in critical and non-critical
many-body systems. More specifically,
compared to the divergence of the block entropy,
we ask the stronger question whether a single spin chain
already contains
an arbitrary amount of entanglement, such that from a 
single specimen a maximally entangled state of arbitrary 
dimension can be distilled.

We will make the argument quantitative by analytically
considering a general framework of translationally invariant 
quantum spin models. As examples in which 
criticality is in one-to-one correspondence with a
divergent single-copy entanglement, we consider 
isotropic spin models, as well as the XY-model. For the 
isotropic XY model we establish the exact asymptotic scaling 
behavior of $\sim (1/6)\log_2 (L)$, and relate it to the
block entropy \cite{Korepin1,Keating}. The results
can also be conceived as statements concerning the 
divergence of fine-grained entanglement \cite{Orus}.

{\it Single-copy entanglement. --}
Let us consider a one-dimensional 
quantum spin system, associated with a Hilbert space
 ${\cal H}= ({\cc}^2)^{\otimes n}$, with a translationally
invariant Hamiltonian.
We distinguish a block of length $L$ of consecutive systems of the
chain. So we have a bi-partioning $n|L$, the
whole system being in a pure state
$\rho=|\psi\rangle\langle \psi| $.

There are several meaninful definitions of single-copy entanglement.
We will primarily be concerned with the question: running a 
physical device 
once, a maximally entangled state of what 
dimension can be distilled from a single
specimen with certainty? Hence, the 
state has a single-copy entanglement
        $E_1(\rho )= \log_2(M)$,
 with respect to the bi-partitioning $n | L$, if $\rho $
 can be deterministically transformed under LOCC
into $|\psi_M\rangle\langle \psi_M|$, i.e., 
a maximally entangled state with
state vector 
$|\psi_M\rangle = (|1,1\rangle +...+ |M,M\rangle)/\sqrt{M}$, so if
 \begin{equation}\label{Single}
       \rho
        \longrightarrow
        |\psi_M\rangle\langle \psi_M| \,\,\text{ under LOCC}.
 \end{equation}
 This is the non-asymptotic analogue of the entropy of
 entanglement of the reduction associated with a block of length $L$.
 Denote with $\alpha_1^\downarrow,...,\alpha_{2^L}^\downarrow$
 the non-increasingly ordered eigenvalues of the reduced state
 with respect to a block of length $L$, then Eq.\ (\ref{Single})
 holds true if and only if \cite{Nielsen}
        $ \sum_{k=1}^K \alpha_k^\downarrow
        \leq
	K/M$ 
 for all $1\leq K\leq M$, so obviously, if and only if
$\alpha_1^\downarrow\leq 1/M$.
  In other words, the transformation is possible
 if the reduction is more mixed in the sense of majorization 
than the
reduction of the maximally entangled state of dimension
 $M\times M$. Given $\alpha^\downarrow_1$,
 the single-copy entanglement is nothing but
        $E_1(\rho ) = \log_2( \lfloor    (\alpha^\downarrow_1 )^{-1} 
         \rfloor)$.
A variant is one allowing for probabilistic 
protocols. For a state $\rho$ we say that
 \begin{eqnarray*}
       E_{\text{p}}(\rho )= \sup \sum_{k=0}^\infty p_k  \log_2(M_k) ,
 \end{eqnarray*}
such that
$\rho$ can be transformed under LOCC into the ensemble
$\{(p_k, |\psi_{M_k}\rangle\langle\psi_{M_k}| ):k=0,1,...\}$. 
This is the average entanglement that can be distilled, allowing
for maximally entangled states of different dimension with 
certain probabilities \cite{Local}. This 
rate is then the solution of a linear program \cite{Dan}.
By definition, we have that $E_1(\rho)\leq E_{\text{p}}(\rho 
)\leq S(\text{tr}_{n\backslash L}[\rho])$, 
where the last inequality follows from the 
fact that 
the entropy of a reduction bounds the rate of 
any (asymptotic) distillation 
protocol. 

Finally, note that for the single-copy
entanglement, superselection rules (SSR)
play a crucial role, notably
a SSR with respect to particle number conservation.
In the presence of SSR, $E_1^{\text{SSR}}$ has to be understood
as referring to a probabilistic transformation distilling
entanglement-EPR states with state vector
$|\psi^{\text{SSR}} \rangle =
(|0,1\rangle|1,0\rangle + |1,0\rangle|0,1\rangle)/\sqrt{2}$
under LOCC and SSR. We will now consider the 
behavior of
the single-copy entanglement in the limit
of large $L$ for critical and non-critical spin chains.

{\it Single-copy entanglement in general 
quantum spin chains. --}
We start from the general 
set of translationally invariant 
quantum spin systems that is as in Ref.\ \cite{Keating} 
mapped onto a fermionic quadratic Hamiltonian under a Jordan-Wigner 
transformation. 
This model embodies a large class of spin models, including the 
anisotropic and isotropic XY-models as important special cases.
Hence, the Hamiltonian is 
\begin{eqnarray}\nonumber
	H= \sum_{q=0,1}
	\sum_{k,j=1}^n
	\left(\frac{B_{j-k}}{2}-\frac{A_{j-k}}
	{4}\right)
	\hat{\sigma}_k^{q+1}
	\biggl[
	\prod_{i\le k+q\atop l\leq j+1-q}
	\hat{\sigma}_i^3
	\hat{\sigma}_l^3\biggr]
	\hat{\sigma}_j^{q+1}
\end{eqnarray}
where $\hat \sigma^1_k, \hat \sigma^2_k,\hat \sigma^3_k$
denote the Pauli operators
associated with site $k=1,...,n$,
equivalent with the fermionic Hamiltonian
\begin{equation}\nonumber
	H=\sum_{j,k=1}^{n}
	\Bigl[
	\hat a_j^\dagger A_{j-k} \hat a_k 
	+ \hat a_j^\dagger B_{j-k} \hat a_k^\dagger
	- \hat a_j B_{j-k} \hat a_k
	\Bigr]
%	- 2\sum_{j=1}^{n}
%	\hat a_j^\dagger \hat a_j.
\end{equation}
The fermionic operators
obey $\{\hat a_j, \hat a_k\}= 0$ and
$\{\hat a_j^\dagger, \hat a_k\}= \delta_{j,k}$. 
The Hamiltonians are related via 
a Jordan-Wigner transformation leading to 
the Hermitian Majorana operators,
        $\hat m_{2i-1} = (\prod_{j<i} \hat \sigma_j^3) \hat \sigma_i^1$ and
        $\hat m_{2i} = (\prod_{j<i} \hat \sigma_j^3) \hat \sigma_i^2$,
where
$\hat a_j = (\hat m_{2j-1} - i \hat 
m_{2j})/2$.
Translational invariance, periodic 
boundary conditions, and Hermicity
are inherited by 
$A_{j},B_j\in\rr$ satisfying
$A_{j}= A_{-j}$, and $B_{j}=-B_{-j}$ for $j=1-n,...,n-1$.
For simplicity, we assume that
there exists a $w\in \nn$ such that $A_j=B_j=0$ for $j>w$.
This model will be our starting point. For all isotropic instances, and 
also for the full XY model we will be able to identify when the 
single-copy entanglement is indeed logarithmically divergent. 
Then, one may distill a maximally entangled state of any
dimension from a single specimen of the 
chain with certainty, 
containing in this sense an ``infinite single-copy entanglement'' 
\cite{Infi}.
We will make use of the powerful methods of Toeplitz determinants 
\cite{Korepin1,Keating,Lieb}. This path is yet in our
instance complicated by the fact that we do not only consider
isotropic models, and that in contrast to the block entropy 
the largest eigenvalue cannot straightforwardly 
be expressed as an integral of a Toeplitz determinant.
The starting point, yet, is the familiar one for
assessing spin systems: 
The ground state of this system is a fermionic 
Gaussian, i.e., quasi-free, state and is 
completely specified by the second moments of 
the Majorana operators. These operators satisfy
$\hat m_j = \hat m_j^\dagger $ and
$\{\hat m_j,\hat m_k\}= 
2 \delta_{j,k}$. The second moments 
can be collected in a 
correlation matrix $\gamma\in \rr^{2n\times 2n}$,
%defined as
%\begin{equation}
	$\text{tr}[\rho \hat m_j \hat m_k]  = \delta_{j,k} + i \gamma_{j,k}$.
%\end{equation}
This matrix is skew-symmetric. 
The entanglement properties of the block of length $L$ can now be inferred
from a principal submatrix $\gamma_L\in\rr^{2L\times 2L}$
of the correlation matrix $\gamma$.  
We consider the entries of $\gamma_L$ in the limit
 of an infinite chain $n\rightarrow\infty$. 
Then, $\gamma_L$ is a block Toeplitz matrix, 
the $l$-th row, $l=1,...,L$,
being given by $(M_{l-1}, M_{l-2},...,M_0,...,M_{l-L})$, with 
$2\times 2$-blocks $M_{L-1},...,M_{1-L}$ that are found to be
\begin{equation}\nonumber
	M_l
	= \left[
	\begin{array}{cc}
	0 & t_l \\
	-t_{-l} & 0 \\
	\end{array}
	\right] , 
	\,\,
	t_l = \frac{1}{2\pi}
	\int_{0}^{2\pi}
%	\frac{g(k)}{|g(k)|}
	g(k)
	e^{-ilk} dk	.
\end{equation}
For no anisotropy, i.e., $B_j=0$ for all $j$, this matrix is a
tensor product of a symmetric matrix and a unit 
skew-symmetric one. 
In generality, we have 
for this model, $g(k):= 
\Lambda(k)/|\Lambda(k)|$,
$\Lambda(k):=A_0+2\sum_{j=1}^{w} A_j\cos(jk)-4i\sum_{j=1}^w 
B_j\sin(jk)$. 
This matrix $\gamma_L$ 
can be brought into a standard normal form $\Gamma_L$
        of a skew-symmetric matrix 
        with an 
        $O\in O(2L)$  preserving the 
	anticommutation relations,
        \begin{equation}\label{Diagonalizing}
                \Gamma_L = O\gamma_L O^T,\,\,	
		\Gamma_L=\bigoplus_{l=1}^L
                \left[
                        \begin{array}{cc}
                        0 & \mu_l\\
                        -\mu_{l} & 0 \\
                        \end{array}
                \right].\nonumber
        \end{equation}
        This defines the quantities $\mu_1 ,...,
        \mu_L \in [0,1]$. 
Such normal mode decompositions have
been employed  
both to evaluate correlation functions \cite{Lieb} and
the block entropy \cite{Latorre1,Korepin1,Keating}.
From now on we will be concerned with the largest eigenvalue
of the reduction of a block of length $L$. 
All eigenvalues $\alpha_1^\downarrow,..., \alpha_{2L}^\downarrow$ 
of the reduction are given by
        $\{\alpha_1^\downarrow,..., \alpha_{2L}^\downarrow\}
	 =\{
        \prod_{l=1}^L
         (1 \pm \mu_l)/2 \}$.
We will be looking at the behavior of the largest eigenvalue $
\alpha_1^\downarrow$ for large $L$. 
This largest eigenvalue is given by
 $
        \alpha_1^\downarrow
        =
        \prod_{l=1}^L (1/2+  \mu_l  /2 )$, or
\begin{equation}\nonumber
        \alpha_1^\downarrow
        =
        \det [  (\id_L +| T_L| )/2 ],
 \end{equation}
$| T_L|  = (T_L^T T_L)^{1/2}$,
where $T_L$ is the 
$L\times L$ 
Toeplitz matrix, with $l$-th row being given by
$(t_{-l+1}, t_{-l+2},...,t_0,...,t_{L-l})$.
The numbers $\mu_1,...,\mu_L$ are the 
singular values of $T_L$.
This matrix $T_L$, satisfying $|T_L|\leq \id_L$, 
is generally not symmetric, as a consequence of 
the anisotropy of the model. Moreover, in contrast to the matrix
$T_L$ itself, $\id_L+  |T_L|$ is not Toeplitz.
In order to show that the single-copy entanglement is 
logarithmically divergent, 
we will make use of appropriate 
bounds that retain this property:
whenever 
the  $A_0,...,A_w$, $B_0,...,B_w$ are such that one can 
prove that the sequence of $L\times L$-Toeplitz matrices
$T_L$ satisfies 
\begin{equation}\label{Start}
	- \log\, | \det[  T_{L}  ]|=\Omega(\log(L))
\end{equation}
(using Landau notation \cite{Landau})
using a Fisher-Hartwig-statement 
\cite{Korepin1,Keating,Lieb}, 
then one can indeed 
conclude that $E_1=\Omega(\log(L))$, i.e,  
the single-copy entanglement diverges at least
logarithmically with increasing block length $L$.
This follows from the following chain,
\begin{eqnarray}\nonumber
%	- \log \alpha^\downarrow_1 &=&
	&-& \log\det[(\id_L + |T_L|)/2]  
	\geq - \frac{1}{2} \log\det[(\id_L +  T_L^T T_L)/2]\nonumber\\
	&\geq & - \frac{1}{4} \log\det[   T_L^T T_L ]=
	-  \frac{1}{2} \log|\det[   T_L ]|\nonumber
\end{eqnarray}
\cite{DetRemark},
where we also have made use of the concavity of the logarithm. 
So, whenever Eq.\ (\ref{Start}) holds,
for appropriate length of the block $L$, 
a maximally entangled pair of any dimension 
can be distilled from a single specimen of the spin chain. 

{\it Isotropic models. --} This case of $B_0,...,B_w=0$
is particularly transparent. 
Here, the asymptotics in $L$ of the determinants 
$\det[M_{x,L}]$ of 
the $L\times L$-Toeplitz matrices 
$M_{x,L}:= ix\id+ (1-x^2)^{1/2}T_L$ 
is known for all $x\in(0,1)$, 
using a Fisher-Hartwig statement. This small detour to
infer about $-\log|\det[T_L]|$  -- corresponding to the case $x=0$ --
is needed as
the Fisher-Hartwig-conjecture has not been proven yet for this 
case. In general, 
one can identify the asymptotic behavior of
determinants of Toeplitz matrices 
by investigating the so-called symbol, see footnote \cite{Symbol}. 
The symbol associated with the Toeplitz matrices $ M_{x,L}$ 
is given by 
\begin{equation}
  G_x(k)= ix + (1-x^2)^{1/2} g(k),\nonumber
\end{equation} 
with $g$ as defined above.
For this class of isotropic models, an explicit factorization of 
the symbol is known \cite{Keating}, see footnote \cite{Thisclass}.
It follows hence from proven instances 
of the Fisher-Hartwig conjecture
that there exists a $c>0$ and an $x_0\in(0,1)$
such that \cite{Landau}
\begin{equation}\nonumber	
 \log | \det[  M_{x,L}  ]|=  c_x \log(L) +o(\log(L))
\end{equation}
with $c_x>c$ for all $x\in(0,x_0)$, whenever 
the function $g$ is discontinuous in $[0,2\pi]$, where
the jumps reflect the Fermi surface. 
From this -- and 
using that $T_L$ has real eigenvalues -- it follows that the
system has a logarithmically divergent single-copy entanglement
if the system is critical \cite{Gap}.
For example, for  the XX model 
this analysis immediately delivers a 
logarithmically divergent single-copy entanglement, whenever
the system is critical.

{\it Anisotropic XY-model. --} For the  XY-model we can conclude
that the single-copy entanglement is
logarithmically divergent if and only if the system is critical.
For this model, we have that $A_0=-1, A_1 = a/2 $ 
and $B_{-1} = - B_{1}= \gamma a /4$, and $0$
elsewhere. For $\gamma=0$, we obtain the XX-model 
(the isotropic XY-model), 
for $a=1$, $\gamma=1$ the critical Ising model. Along the line 
$\gamma\in[-1,1]$, $a=1$ the anisotropic 
model is critical. Then, we encounter a generally 
non-symmetric matrix 
$T_L$. The associated symbol is given by
\begin{eqnarray}\nonumber
	g(k)
	= \frac{a \cos(k) - 1 + i a \gamma \sin(k) }
	{
	((a\cos(k)-1)^2 + \gamma^2 a^2 \sin^2 (k))^{1/2}}.
\end{eqnarray}
For $\gamma\neq 0$ and 
$1/ a 
\in (0,1)$, the symbol is continuous, and one
finds a saturating block entropy \cite{Korepin1}
(and hence a saturating single-copy
entanglement). 
Along the critical line $a=1$, $\gamma\in(-1,1)$, 
in turn, we can identify the explicit factorization of the
discontinuous symbol. 
There is a single discontinuity at $k_1=0$
\cite{BoundaryNote},
and in the terms of footnote \cite{Thisclass}
we find $\beta_1 = 1/2$, so that $g(k)$ can be decomposed as
\begin{equation}\nonumber
  g(k) = \phi(k) t_{1/2}(k), 
\end{equation}
where $\phi$ is a continuously differentiable
function. For the case of a single discontinuity and $\alpha_1=0$, 
the Fisher-Hartwig conjecture has been
proven for any $\beta_1\in \cc$ with $ \Re(\beta_1) <5/2$ \cite{Libby},
including our case at hand. 
Hence, we find $-\log |\det[T_L]|= \Omega(\log(L))$, and
hence $E_1= \Omega(\log(L))$. 
Together with the result of the subsequent section this shows that the
single-copy entanglement of the XY-model is logarithmically divergent 
exactly if the model is critical. Note that this implies also a less technical
alternative proof of the logarithmic 
divergence of the block entropy in the
critical XY model.

{\it Scaling of single-copy entanglement in the XX-model. --}
In the light of these findings, it is interesting to 
see how the exact asymptotic behavior is compared to that
of the block entropy, including prefactors. 
We make this specific
for the isotropic XX-model, 
where now $T_L = T_L^T$.
The technicality when evaluating $
\alpha_1^\downarrow  = \det[
(\id_L + |T_L|)/2]$ that we encounter
here is that the function 
$f:\cc\rightarrow\cc$, 
$f(x) :=  \log_2((1+|x|)/2)$, is not analytic.
So before we can exploit Fisher-Hartwig-type results, 
we have to approximate
$\alpha_1^\downarrow $ with sequences
based on functions with appropriate
continuity properties. We can take any functions
$f_*:\cc\times \rr_+ \rightarrow\cc$
which are analytic on $\{z\in \cc: \Im(z)<\delta\}$
for a $\delta>0$, such that 
on the real axis $\lim_{\delta\searrow 0} f_*(x,\delta)=
f(x,0)$ for $x\in\rr$. Take, e.g.,
\begin{equation}\nonumber
f_*(z,\delta):= \log(1/2 + 
(z^2 + \delta^2)^{1/2}/2).
\end{equation}

We are then in the position to identify the 
asymptotic behavior of the single-copy entanglement.
This can be done similarly to 
Ref.\ \cite{Korepin1} using 
the characteristic polynomial 
$F: \cc\rightarrow \cc$ of $T_L$ defined as
$ F(\lambda):=  \det[\lambda \id_L - T_L]$:
the function $F$ is meromorphic, and all zeros are in
the interval $[-1,1]$. One can hence write
\begin{eqnarray}\label{F}
	d_*  =
	\lim_{\delta\searrow 0}
	\lim_{\varepsilon\searrow 0}
	\frac{1}{2\pi i}
	\int  dz 
	f_*
	(z,\delta)
	\frac{F'(z)}{F(z)}
\end{eqnarray}
where the integration path is chosen to enclose
the interval $[-1,1]$, with path
from $(-1-\delta+i\varepsilon, 1+\delta + i\varepsilon)$, 
towards the negative real numbers along a 
circle segment with radius $\delta/2$,  
then $(1+\delta -i \varepsilon, -1-\delta- 
i\varepsilon)$, and again
along a circle segment to $-1-\delta+i\varepsilon$, such that
$\lim_{\delta\searrow 0}d_*=d$. 
The symbol of $\lambda\id - T_L$
with factorization as in Eq.\ (\ref{Factor})
for the XX-model is known \cite{Korepin1}, see footnote 
\cite{Thisclass}.
Using  a  Fisher-Hartwig statement, we find that the linear
terms in $L$ do not contribute, using Cauchy's 
theorem and using that 
$\lim_{\delta\searrow 0}f_*(\pm1,\delta)=0$,
and finally arrive at
\begin{eqnarray}\nonumber
	d  &=& \log(L) \frac{2}{\pi^2}
	\int_{-1}^1 \frac{\log_2[(1+|x|)/2]}{1-x^2} dx  + 
	o(\log(L)).
\end{eqnarray}
This in turn finally implies that whenever $1/a \in[-1,1]$ and
the XX-model is critical, we observe the scaling behavior
\begin{eqnarray}\nonumber
        E_1  =  
        \frac{1}{6 } \log_2 (L) + o(\log(L))  ,
\end{eqnarray}
independent of $a$; it saturates in the 
non-critical case. This result is astonishing:
the single-copy entanglement does not only diverge, 
but  has up to a factor of two the same asymptotic behavior 
as the entropy of entanglement scaling as 
$S = (1/3) \log_2 (L) + o(\log(L))$. 
Half of the asymptotically
distillable entanglement is hence already available 
on the single-shot level.
 
{\it Outlook and summary. --}
Finally, let us comment on the crucial 
role of SSR for
the single-copy entanglement. This 
is relevant, e.g., when assessing the single-copy entanglement
in the hard-core limit of the Bose-Hubbard model (infinite repulsion 
energy) \cite{Sachdev}. There, the Hamiltonian is isomorphic to the XX-model, 
via the mapping $\hat \sigma_j^1 = \hat b_j + \hat b_j^\dagger$,
$\hat \sigma_j^2 = -i ( \hat b_j - \hat b_j^\dagger)$, and
$\hat \sigma_j^3 = 1 - 2 \hat b_j^\dagger \hat b_j$ for each 
site $j$.
Yet, the concept of entanglement is different due 
to the presence of a particle number conservation 
SSR in the former case: 
Transformations under LOCC have to be replaced
by those under LOCC+SSR. The single-copy
entanglement in the above sense can however still be
efficiently evaluated in $L$; and these superselection 
rules must be respected when assessing single-copy 
entanglement.

In this paper, we have fleshed out
the notion of single-copy entanglement in quantum spin chains. 
Such a notion is the appropriate one when one is not interested 
in the entanglement properties of an asymptotic supply of a
identically prepared many-body systems, but of single specimens. 
It is the hope that these findings also serve as a guideline when
assessing entanglement in actual experimental situations, let
it be in condensed-matter systems or in systems of ultracold atoms 
in optical lattices.

{\it Acknowledgements. --}
We would like to thank 
J.I.\ Cirac,
T.\ Cubitt, 
M.B.\ Plenio,
D.\ Schlingemann, 
R.F.\ Werner, and
M.M.\ Wolf for discussions.
This work has been supported by the DFG
(SPP 1116, SPP 1078),
the EU (QUPRODIS), the EPSRC, and the
European Research Councils (EURYI).


\begin{thebibliography}{99}
\bibitem{Landau}
	For functions of
	$L$, $f= O(g)$ means that $f\leq c g$ for some $c>0$ 
	and all $L$, 
	$f= o(g)$ means that $f/g\rightarrow 0$ for $L\rightarrow\infty$,
	and $f= \Omega(g)$ that  $f\geq C g$ for some $C>0$ 
	and all $L$.
\bibitem{Symbol}        
        	For an square summable sequence $\{ s_k \}_k$, $s_k\in\cc$,
	the symbol $G$
	corresponding to a sequence of $L\times L$ Toeplitz
	matrices with $l$-th row
	$(s_{l-1},...,s_{L-l})$ is given by
	$        G(k)= \sum_{j=-\infty}^\infty e^{ij k} s_j$.
\bibitem{Thisclass}
	For this class of isotropic models, 
	an explicit factorization of 
	the symbol is given by
	\begin{eqnarray}\nonumber
	G_x(k) & =&  - (1-x^2)^{1/2} 
	h(x)^{\sum_{r=1}^R (-1)^r k_r /\pi}\nonumber\\
	&\times &
	\prod_{r=1}^R t_{(-1)^r f(x) }(k- k_r)
	t_{-(-1)^r f(x) } (k+k_r),\label{Factor}
	\end{eqnarray}	
	$t_\beta(k):=\exp(-i\beta (\pi - k\mod 2\pi))$,
	$h(x) := (i x - (1-x^2)^{1/2} )/( i x + (1-x^2)^{1/2})$
	and $f(x) :=  (\log\circ h)(x)/(2\pi i)$. 
	$k_r$, $r=1,...,R$, are the 
		$k$ for which $g$ is discontinuous.
\bibitem{Gap}
	Here, we use critical in the sense that there
	is no gap between the energy of the ground 
	and the first excited state.
\bibitem{Infi}
	The notion of entanglement in systems with 
	infinitely many degrees of freedom can also be
	grasped mathematically sharply in 
	the language of von-Neumann algebras 
	\cite{InfinitelyEntangled}.
\bibitem{BoundaryNote}
	Here, $k_1=0$, which is at the boundary of the 
	domain of $g$. Then the integral has to be
	understood as ranging from $-\varepsilon, 
	2\pi-\varepsilon$, $\varepsilon>0$, 
	see Refs.\ \cite{Libby}.
\end{thebibliography}
\end{document}